\RequirePackage{fix-cm}
\documentclass[twocolumn,epjc3]{svjour3}  
\smartqed  

\RequirePackage{graphicx}
\RequirePackage[all]{xy}
\RequirePackage{amsmath}
\RequirePackage{amssymb}
\RequirePackage{epstopdf}
\RequirePackage[colorlinks=true, pdfstartview=FitV, linkcolor=blue, citecolor=red, urlcolor=magenta]{hyperref}
\RequirePackage{latexsym}
\RequirePackage{amsfonts}
\RequirePackage{subeqnarray}
\RequirePackage{widetext}
\newcommand{\be}{\begin{equation}}
\newcommand{\ee}{\end{equation}}
\newcommand{\ben}{\begin{eqnarray}}
\newcommand{\een}{\end{eqnarray}}
\newcommand{\bes}{\begin{subequations}}
	\newcommand{\ees}{\end{subequations}}
\newcommand{\bn}{\begin{eqnarray}}
\newcommand{\en}{\end{eqnarray}}

\newcommand{\bb}{\bibitem}
\journalname{Eur. Phys. J. C}

\begin{document}
	
	\title{Building analytical three-field cosmological models}
	
	
	\author{J. R. L. Santos \thanksref{e1,addr1}
		\and P. H. R. S. Moraes\thanksref{e2,addr2}
		\and
		D. A. Ferreira \thanksref{e3,addr1,addr3}
		\and
		D. C. Vilar Neta\thanksref{e4,addr1,addr4}        
	}

	\thankstext{e1}{e-mail: joaorafael@df.ufcg.edu.br}
	\thankstext{e2}{e-mail: moraes.phrs@gmail.com}
	\thankstext{e3}{e-mail: dalvesferreira1991@hotmail.com}
	\thankstext{e4}{e-mail: deusaletecamara@yahoo.com.br}

	
	\institute{Unidade Acad\^emica de F\'isica, Universidade de Federal de Campina Grande, 58109-970 Campina Grande, PB, Brazil.\label{addr1}
		\and 
		ITA - Instituto Tecnol\'ogico de Aeron\'autica, 12228-900 S\~ao Jos\'e dos Campos, SP, Brazil.\label{addr2}
		\and
		Departamento de F\'{i}sica, Universidade Federal da Para\'{i}ba, 58051-970 Jo\~ao Pessoa, PB, Brazil. \label{addr3}
		\and  
		Departamento de F\'{i}sica, Universidade Estadual da Para\'{i}ba, 58429-500 Campina Grande, PB, Brazil.\label{addr4}
	}
	
	\date{Received: date / Accepted: date}

	\maketitle
	
	\begin{abstract}
		A difficult task to deal with is the analytical treatment of models composed by three real scalar fields, once their equations of motion are in general coupled and hard to be integrated. In order to overcome this problem we introduce a methodology to construct three-field models based on the so-called ``extension method''. The fundamental idea of the procedure is to combine three one-field systems in a non-trivial way, to construct an effective three scalar field model. An interesting scenario where the method can be implemented is within inflationary models, where the Einstein-Hilbert Lagrangian is coupled with the scalar field Lagrangian. We exemplify how a new model constructed from our method can lead to non-trivial behaviors for cosmological parameters. 
	\end{abstract}

\section{Introduction}
\label{sec_1}

Since 1970's, topological defects have been investigated as promising analytical solutions in high energy physics and in ferromagnet models \cite{rajaraman}-\cite{montonen}. In the last decades, these defects were applied in several different scenarios, like braneworld models, condensate matter, besides Einstein-Hilbert and generalized cosmology \cite{trullinger},\cite{blochbranes}-\cite{ms_16}. This applicability growth  was accompanied by the emergence of new mathematical methods to treat topological defects, especially when we talk about models composed by two or more scalar fields. 

A relevant methodology which should be highlighted is the so-called BPS (Bogolmon'y-Prasad-Sommerfield) method \cite{bps}, that enables one to determine analytical solutions for one or more real scalar field models from first-order differential equations, instead standard second order equations of motion. Moreover, BPS solutions, or BPS states, are associated with the minimal energetic solutions of static physical systems. However, when we deal with Lagrangian densities composed by two or more real fields, even the BPS first-order differential equations are very hard to be integrated once they are generally coupled. Thus, we need specific methodologies to find analytical defects for two or more scalar field models. 

An interesting method to solve two-field systems was proposed by Bazeia {\it et al.} \cite{bflr_02}, where inspired by Rajaraman's trial orbit method \cite{rajaraman_79}, the authors introduced an approach to find analytical solutions for the coupled first-order differential equations of such systems. One of the most popular models solved by this method is denominated  ``BNRT'' \cite{bnrt} and it has been applied in several different contexts, as one can see in \cite{blochbranes},\cite{bllm}-\cite{ch_10}. As an alternative for the trial orbit presented in \cite{bflr_02}, de Souza Dutra \cite{dutra_orb} constructed new orbits for the BNRT models through the so-called ``integrating factor method''. Despite the success of such methodologies, the challenge of finding new analytical models formed by one, two or more real scalar fields remains tricky. 

In scenarios composed by one scalar field, new analytical models can be generated with the deformation method proposed by Bazeia, Losano and Malbouisson \cite{blm}. Such a method is based on a connection between two one-field models, via the denominated deformation function. So, if we know the deformation function and an analytical one-field model, we are able to generate several families of new scalar field systems, as one can see in \cite{blm}-\cite{bglms}. 

Inspired by the deformation method, Bazeia, Losano and Santos \cite{bls} introduced the extension method to construct analytical two scalar field models, starting from one-field ones. The advantage of such a methodology is that a possible set of solutions for the equations of motion of the two-field model is exactly formed by the solutions of the one-field systems used in the construction process. The last method was applied in the quintessence cosmology, leading to new sets of analytical cosmological parameters \cite{ms_14}. 

 As a motivation to apply our methodology in the context of cosmological models, we can point that a multi-field inflation is able to yield to a proper relation between the tensor-to-scalar ratio and the spectral index, as pointed by Ellis {\it et al. } \cite{ellis_2014}. Moreover, multiple fields allow us to new features for the physical systems, as well as, for the cosmological parameters, which cannot be derived from single field models, as pointed very recently in \cite{abedi_2017,bjorkmo_2017}.

Our aim in the present work is to increase the amount of analytical three scalar field models inspired by the extension method. We believe that this approach can overcome several difficulties related with the integration process of coupled first-order differential equations. In order to show the applicability of our procedure as well as its robustness, we will use it to build a three-field quintessence model.

 The article is organized as follows: section \ref{sec_2} shows some generalities about the deformation method and about the BPS approach for three scalar field models. In section \ref{sec_3} we present a new version for the extension method while its applicability is discussed carefully in section \ref{sec_4}, where we construct several examples. In section \ref{sec_5}  we establish the bases for our quintessence model, we apply one of our examples in this context and we analyze the cosmological features of the effective model. Final remarks and perspectives of this methodology are reported in section \ref{sec_6}.

\section{Generalities}
\label{sec_2}

We begin our analysis with a review about generalities which are in the foundation of our method. Let us start with the so-called deformation method, firstly presented by Bazeia, Losano and Malbouisson \cite{bls}. This method proposes a connection between two one-field Lagrangian densities, which may have the forms
\be \label{sec2_eq1}
{\cal L}=\frac{1}{2}\,\partial_{\,\mu}\,\phi\,\partial^{\,\mu}\phi-V(\phi)\,; \,\,\, {\cal L}_d=\frac{1}{2}\,\partial_{\,\mu}\,\chi\,\partial^{\,\mu}\chi-U(\chi)\,,
\ee
where $V$ and $U$ are their respective potentials and $\mu=0,1$ if we are working in $1+1$ space-time. The equations of motion for both theories can be derived in a straightforward way, yielding
\be \label{sec2_eq2}
\phi^{\,\prime\prime}=V_{\,\phi}\,; \qquad \chi^{\,\prime\prime}=U_{\,\chi}\,; \,\,\, V_{\,\phi}=\frac{d\,V}{d\,\phi}\,; \,\,\, U_ {\,\chi}=\frac{d\,U}{d\,\chi}\,,
\ee
if we are dealing with static fields, i.e., $\phi=\phi(x)$ and $\chi=\chi(x)$, and with a metric signature $(1,-1)$. Besides, the primes mean derivatives in respect to the $x-$coordinate. 

The previous equations can be integrated once, giving rise to the following first-order differential equations
\be \label{sec2_eq3}
\phi^{\,\prime}=\pm\,\sqrt{2\,V}=\pm\,W_{\phi}(\phi)\,; \,\,\, \chi^{\,\prime}=\pm\,\sqrt{2\,U}=\pm\,\widetilde{W}_{\chi}(\chi)\,,
\ee
where we defined
\be \label{sec2_eq4}
V=\frac{W_{\,\phi}^2}{2}\,; \qquad U=\frac{\widetilde{W}_{\,\chi}^{\,2}}{2}\,,
\ee
with $W_{\phi}=d\,W/d\,\phi$, $\widetilde{W}_{\chi}=d\,\widetilde{W}/d\,\chi$, and $W(\phi)$ and $\widetilde{W}(\chi)$ are called superpotentials. Both scalar fields are mapped if we consider $\phi=f(\chi)$, and $\chi=f^{\,-1}(\phi)$, where $f$ is named ``deformation function''. Therefore, replacing the deformation function in the first-order differential equation for the field $\phi$, we find the constraints
\ben \label{sec2_eq5}
&&
\frac{d\,\phi}{d\,\chi}=\frac{W_{\,\phi}(\phi)}{W_{\,\chi}(\chi)}\,; \\ \nonumber
&&
U(\chi)=\frac{V\,(\phi=f(\chi))}{f_{\chi}^{\,2}}\,; \,\, \widetilde{W}_{\chi}=\frac{W_{\phi}}{f_{\,\chi}}\Bigg|_{\phi=f(\chi)}\,,
\een
with $f_{\chi}=d\,f/d\,\chi$. Consequently, if we know the potential and the solution for the model described by ${\cal L}$, we can use these results together with the deformation function to build the model ${\cal L}_d$, which is the deformed Lagrangian density. 

Let us now review some basic concepts about the first-order formalism for a three scalar field Lagrangian density. Supposing the following action
\ben \label{sec2_eq6}
&&
S=\int\,dt\,dx\,{\cal L} \\ \nonumber
&&
=\int\,dt\,dx\,\left[\sum_{i}\,\frac{1}{2}\,\partial_{\mu}\phi_i\,\partial^{\,\mu}\phi_i-V(\phi_1,\phi_2,\phi_3)\right]\,, 
\een
with $i=1,\,2,\,3$, $\phi_1=\phi$, $\phi_2=\chi$, and $\phi_3=\xi$, we can minimize it to derive the equations of motion
\be \label{sec2_eq7}
\phi^{\,\prime\,\prime}=V_{\phi}\,; \qquad \chi^{\,\prime\,\prime}=V_{\chi}\,; \qquad \xi^{\,\prime\prime}=V_{\xi}\,,
\ee
for static fields. Withal, the total energy for this static model is given by
\be \label{sec2_eq8}
E=-\int\,dx\,{\cal L}=\int\,dx\,\left[\sum_{i}\frac{\phi_i^{\,\prime\,2}}{2}+V(\phi,\chi,\xi)\right]\,,
\ee
where the potential $V$ can be defined as
\be \label{sec2_eq9}
V=\frac{W_{\,\phi}^{\,2}}{2}+\frac{W_{\chi}^{\,2}}{2}+\frac{W_{\,\xi}^{\,2}}{2}\,; \qquad W=W(\phi,\chi,\xi)\,.
\ee
By applying the BPS (Bogolmon'y-Prasad-Sommerfield) method \cite{bps}, we are able to rewrite the total energy as
\ben \label{sec2_eq10}
&&
E=\frac{1}{2}\,\int\,dx\,\bigg[(\phi^{\,\prime} \mp W_{\,\phi})^2+(\chi^{\,\prime} \mp W_{\,\chi})^2 \\ \nonumber
&&
+(\xi^{\,\prime} \mp W_{\,\xi})^2\pm2\,W_{\,\phi}\,\phi^{\,\prime}\pm2\,W_{\,\chi}\,\chi^{\,\prime}\pm2\,W_{\,\xi}\,\xi^{\,\prime}\bigg]\,,
\een
therefore, if the first-order differential equations
\be \label{sec2_eq11}
\phi^{\,\prime}=\pm\,W_{\,\phi}\,; \qquad \chi^{\,\prime}=\pm\,W_{\,\chi}\,; \qquad \xi^{\,\prime}=\pm\,W_{\,\xi}\,,
\ee
are satisfied, we find the total energy
\ben \label{sec2_eq12}
\nonumber
E_{BPS}&=&|E|=\int\,dx\,\left(W_{\phi}\,\phi^{\,\prime}+W_{\,\chi}\,\chi^{\,\prime}+W_{\,\xi}\,\xi^{\,\prime}\right) \\
&
=&\int\,dx\,\frac{d\,W}{d\,x}=\left|\Delta\,W\right|\,,
\een
with 
\be \label{sec2_eq13}
\Delta\,W=W(\infty,\infty,\infty)-W(-\infty,-\infty,-\infty)\,.
\ee

The main difference between Eq. (\ref{sec2_eq3}) and the first-order differential equations shown in (\ref{sec2_eq11}) is that the last are in general coupled and hard to be integrated (that is what makes analytical three-field models so hard to be found). A well-known method to integrate equations presented in (\ref{sec2_eq11}) is the integrating factor, which consists of rewriting them as
\be \label{sec2_eq14}
\frac{d\,\phi}{d\,\chi}=\frac{W_{\,\phi}}{W_{\,\chi}}\,; \qquad \frac{d\,\phi}{d\,\xi}=\frac{W_{\,\phi}}{W_{\,\xi}}\,; \qquad \frac{d\,\xi}{d\,\chi}=\frac{W_{\,\xi}}{W_{\,\chi}}\,,
\ee
with the solutions $\phi=\phi(\chi)$, $\phi=\phi(\xi)$ and $\xi=\xi(\chi)$ being denominated ``orbits''.

	\section{The method}
	\label{sec_3}
	
	The method to construct new analytical three-field models will be a generalization of the extension method for two scalar fields, introduced by Bazeia, Losano and Santos \cite{bls}. In order to establish the three-field version for the extension method, we are going to use two deformation functions to rewrite the left-hand side of equation (\ref{sec2_eq5}) as the first differential equation presented in (\ref{sec2_eq14}). An analogous procedure can be repeated to generate the other two first-order differential equations shown in (\ref{sec2_eq14}).
	
	This mechanism means that we can build a three scalar field model combining three one-field systems. Moreover, once we know the solutions of the three one-field models as well as the deformation functions, the effective three-field model is going to be analytically solvable. Such a nice feature agrees with the results derived in \cite{bls}. 
	
	So, establishing $\phi=f_{1}\,(\chi)$, $\chi=f_1^{\,-1}(\phi)$ and $\phi=f_{2}\,(\xi)$, $\xi=f_2^{\,-1}(\phi)$, we can rewrite $\phi^{\,\prime}=W_{\,\phi}(\phi)$ in seven different but equivalent forms, given by
	
	\ben \label{sec3_eq1}
	&& \nonumber
	\phi^{\prime}=W_{\phi}(\phi)\,;\qquad \phi^{\prime}=W_{\phi}(\chi)\,;\qquad \phi^{\prime}=W_{\phi}(\phi,\chi)\,; \\ \nonumber
	&&
	\phi^{\prime}=W_{\phi}(\xi)\,; \qquad \phi^{\prime}=W_{\phi}(\phi,\xi)\,; \qquad \phi^{\prime}=W_{\phi}(\chi,\xi)\,;\\
	&&
	\phi^{\prime}=W_{\phi}(\phi,\chi,\xi)\,.
	\een
	Analogously, if $\xi=f_3(\chi)$, $\chi=f_3^{\,-1}(\xi)$ where $f_3=f_2^{\,-1}(\phi=f_1)$, the first-order equations $\chi^{\,\prime}=W_{\chi}(\chi)$ and $\xi^{\,\prime}=W_{\,\xi}(\xi)$ are also represented as  
	\bn \label{sec3_eq2}
	&& \nonumber
	\chi^{\prime}=W_{\chi}(\phi)\,;\qquad \chi^{\prime}=W_{\chi}(\chi)\,;\qquad \chi^{\prime}=W_{\chi}(\phi,\chi)\,;\\ \nonumber
	&&
	\chi^{\prime}=W_{\chi}(\xi)\,;\qquad \chi^{\prime}=W_{\chi}(\phi,\xi)\,; \qquad \chi^{\prime}=W_{\chi}(\chi,\xi)\,; \\
	&&
	\chi^{\prime}=W_{\chi}(\phi,\chi,\xi)\,,
	\en
	\bn \label{sec3_eq3}
	&& \nonumber
	\xi^{\prime}=W_{\xi}(\phi)\,;\qquad \xi^{\prime}=W_{\xi}(\chi)\,;\qquad \xi^{\prime}=W_{\xi}(\phi,\chi)\,;  \\ \nonumber
	&&
	\xi^{\prime}=W_{\xi}(\xi)\,; \qquad \xi^{\prime}=W_{\xi}(\phi,\xi)\,; \qquad \xi^{\prime}=W_{\xi}(\chi,\xi)\,; \\
	&&
	\xi^{\prime}=W_{\xi}(\phi,\chi,\xi)\,.
	\en
	
	Therefore, with the previous ingredients in hands it is possible to rewrite $d\,\phi/d\,\chi$ in Eq. (\ref{sec2_eq5}) as the following:
\begin{widetext}
	\bn \label{sec3_eq4}
	\nonumber
	&&
	\frac{d\phi}{d\chi} = \bigg[a_{11}W_{\phi}(\chi)+a_{12}W_{\phi}(\phi,\chi)+a_{13}W_{\phi}(\phi)+a_{14}W_{\phi}(\xi)+a_{15}W_{\phi}(\phi,\xi)+a_{16}W_{\phi}(\chi,\xi)
	+a_{17}W_{\phi}(\phi,\chi,\xi) \\ \nonumber
	&&
	+c_{11}g(\chi)+c_{12}g(\phi,\chi)+c_{13}g(\phi)+c_{14}g(\xi)+c_{15}g(\phi,\xi)+c_{16}g(\chi,\xi)
	+c_{17}g(\phi,\chi,\xi)\bigg]\,  \,\bigg[b_{1}W_{\chi}(\chi) +b_{2}W_{\chi}(\phi,\chi) \\ \nonumber
	&&
	+b_{3}W_{\chi}(\phi)+b_{4}W_{\chi}(\xi)+b_{5}W_{\chi}(\phi,\xi)+b_{6}W_{\chi}(\chi,\xi)+b_{7}W_{\chi}(\phi,\chi,\xi)+c_{21}\tilde{f}(\chi)+c_{22}\tilde{f}(\phi,\chi) +c_{23}\tilde{f}(\phi)+c_{24}\tilde{f}(\xi) \\ 
	&&
	+c_{25}\tilde{f}(\phi,\xi)+c_{26}\tilde{f}(\chi,\xi)+c_{27}\tilde{f}(\phi,\chi,\xi)\bigg]^{-1}=\frac{W_\phi}{W_\chi},
	\en
\end{widetext}
	where $a_{ij}$, $b_{j}$, $c_{ij}$ for $i=1,2$, and $j=1,2,3,4,5,6,7$ are real constants which obey the constraints $a_{11}+a_{12}+a_{13}+a_{14}+a_{15}+a_{16}+a_{17}=1$, $b_{1}+ b_{2}+ b_{3}+ b_{4}+ b_{5}+ b_{6}+ b_{7}=1$, $c_{11}+ c_{12}+ c_{13}+ c_{14}+ c_{15}+ c_{16}+ c_{17}=0$ and $c_{21}+ c_{22}+ c_{23}+ c_{24}+ c_{25}+ c_{26}+ c_{27}=0$. We also have $W_\phi$ and $W_\chi$ without specific functional dependence representing the derivatives of an effective three-field superpotential $W=W(\phi,\chi,\xi)$ in respect to $\phi$ and $\chi$, respectively. Then, we see that Eq. (\ref{sec3_eq4}) has the same form as the left-hand side expression in (\ref{sec2_eq14}).
	
	Moreover, the functions $ \tilde{f}$ and $g$ are responsible for connecting the fields $\phi$ and $\chi$ in this effective three-field model via the additional constraint
	\be \label{sec3_eq5}
	W_{\phi\chi}=W_{\chi\phi}\,,
	\ee
	thus, using Eq. $(\ref{sec3_eq4})$ in this last equation leads to
	\bn \label{sec3_eq6}
	&&
	\nonumber
	a_{11}W_{\phi\chi}(\chi)+a_{12}W_{\phi\chi}(\phi,\chi)+a_{16}W_{\phi\chi}(\chi,\xi) \\ \nonumber
	&&
	+a_{17}W_{\phi\chi}(\phi,\chi,\xi)+c_{11}g_{\chi}(\chi)+c_{12}g_{\chi}(\phi,\chi)
	 \\ \nonumber
	&&
	+c_{16}g_{\chi}(\chi,\xi)+c_{17}g_{\chi}(\phi,\chi,\xi) = b_{2}W_{\chi\phi}(\phi,\chi) \\ 
	&&
	+b_{3}W_{\chi\phi}(\phi)+b_{5}W_{\chi\phi}(\phi,\xi)+b_{7}W_{\chi\phi}(\phi,\chi,\xi) \\ \nonumber
	&&
	+c_{22}\tilde{f}_{\phi}(\phi,\chi) +c_{23}\tilde{f}_{\phi}(\phi)+c_{25}\tilde{f}_{\phi}(\phi,\xi)+c_{27}\tilde{f}_{\phi}(\phi,\chi,\xi).
	\en
	This previous procedure can be repeated for $d\,\xi/d\,\chi$ as well as for $d\,\phi/d\,\xi$, yielding 
\begin{widetext}
	\bn \label{sec3_eq7}
	&&
	\nonumber
	\frac{d\xi}{d\chi} = \bigg[a_{21}W_{\xi}(\chi)+a_{22}W_{\xi}(\phi,\chi)+a_{23}W_{\xi}(\phi)+a_{24}W_{\xi}(\xi)+a_{25}W_{\xi}(\phi,\xi)+a_{26}W_{\xi}(\xi,\chi)+a_{27}W_{\xi}(\phi,\chi,\xi) \\ \nonumber
	&&
	+c_{31}\tilde{g}(\chi)+c_{32}\tilde{g}(\phi,\chi)+c_{33}\tilde{g}(\phi)+c_{34}\tilde{g}(\xi)+c_{35}\tilde{g}(\phi,\xi)+c_{36}\tilde{g}(\chi,\xi)+c_{37}\tilde{g}(\phi,\chi,\xi)\bigg]\, \times \,\bigg[b_{1}W_{\chi}(\chi) +b_{2}W_{\chi}(\phi,\chi) \\ \nonumber
	&&
	+b_{3}W_{\chi}(\phi)+b_{4}W_{\chi}(\xi)+b_{5}W_{\chi}(\phi,\xi)+b_{6}W_{\chi}(\chi,\xi)+b_{7}W_{\chi}(\phi,\chi,\xi)+c_{21}\tilde{f}(\chi)+c_{22}\tilde{f}(\phi,\chi) +c_{23}\tilde{f}(\phi)+c_{24}\tilde{f}(\xi) \\ 
	&&
	+c_{25}\tilde{f}(\phi,\xi)+c_{26}\tilde{f}(\chi,\xi)+c_{27}\tilde{f}(\phi,\chi,\xi)\bigg]^{-1}=\frac{W_{\xi}}{W_{\chi}},
	\en
	with $\tilde{g}$ as another connection function, $a_{21}+a_{22}+a_{23}+a_{24}+a_{25}+a_{26}+a_{27}=1$, $c_{31}+ c_{32}+ c_{33}+ c_{34}+ c_{35}+ c_{36}+ c_{37}=0$ and
	\bn \label{sec3_eq8}
	&&
	\nonumber
	\frac{d\phi}{d\xi} = \bigg[a_{11}W_{\phi}(\chi)+a_{12}W_{\phi}(\phi,\chi)+a_{13}W_{\phi}(\phi)+a_{14}W_{\phi}(\xi)+a_{15}W_{\phi}(\phi,\xi)+a_{16}W_{\phi}(\chi,\xi)+a_{17}W_{\phi}(\phi,\chi,\xi) \\ \nonumber
	&&
	+c_{11}g(\chi)+c_{12}g(\phi,\chi)+c_{13}g(\phi)+c_{14}g(\xi)+c_{15}g(\phi,\xi)+c_{16}g(\chi,\xi)+c_{17}g(\phi,\chi,\xi)\bigg]\, \times \,\bigg[a_{21}W_{\xi}(\chi)+a_{22}W_{\xi}(\phi,\chi) \\ \nonumber
	&&
	+a_{23}W_{\xi}(\phi)+a_{24}W_{\xi}(\xi)+a_{25}W_{\xi}(\phi,\xi)+a_{26}W_{\xi}(\xi,\chi)+a_{27}W_{\xi}(\phi,\chi,\xi)+c_{31}\tilde{g}(\chi)+c_{32}\tilde{g}(\phi,\chi)+c_{33}\tilde{g}(\phi)+c_{34}\tilde{g}(\xi) \\ 
	&&
	+c_{35}\tilde{g}(\phi,\xi)+c_{36}\tilde{g}(\chi,\xi)+c_{37}\tilde{g}(\phi,\chi,\xi)\bigg]^{-1}=\frac{W_\phi}{W_\xi}.
	\en
\end{widetext}
	As in the first application of the extension method, equations (\ref{sec3_eq7}) and (\ref{sec3_eq8}) impose the extra constraints
	\be\label{sec3_eq9}
	W_{\xi\chi}=W_{\chi\xi};
	\ee
	\bn\label{sec3_eq10}
	&&
	\nonumber
	a_{21}W_{\xi\chi}(\chi)+a_{22}W_{\xi\chi}(\phi,\chi)+a_{26}W_{\xi\chi}(\xi,\chi) \\ \nonumber
	&&
	+a_{27}W_{\xi\chi}(\phi,\chi,\xi)+c_{31}\tilde{g}_{\chi}(\chi)+c_{32}\tilde{g}_{\chi}(\phi,\chi) \\ \nonumber
	&&
	+c_{36}\tilde{g}_{\chi}(\chi,\xi)+c_{37}\tilde{g}_{\chi}(\phi,\chi,\xi)=b_{4}W_{\chi\xi}(\xi) \\ 
	&&  
	+b_{5}W_{\chi\xi}(\phi,\xi)+b_{6}W_{\chi\xi}(\chi,\xi)+b_{7}W_{\chi\xi}(\phi,\chi,\xi) \\ \nonumber
	&&
	+c_{24}\tilde{f}_{\xi}(\xi)+c_{25}\tilde{f}_{\xi}(\phi,\xi)+c_{26}\tilde{f}_{\xi}(\chi,\xi)+c_{27}\tilde{f}_{\xi}(\phi,\chi,\xi)\,,
	\en
	and
	\be \label{sec3_eq11}
	W_{\phi\xi}=W_{\xi\phi}\,;
	\ee
	\bn \label{sec3_eq12}
	&&
	\nonumber
	a_{14}W_{\phi\xi}(\xi)+a_{15}W_{\phi\xi}(\phi,\xi)+a_{16}W_{\phi\xi}(\chi,\xi) \\ \nonumber
	&&
	+a_{17}W_{\phi\xi}(\phi,\chi,\xi)+c_{14}g_{\xi}(\xi)+c_{15}g_{\xi}(\phi,\xi)\\
	&&
	\nonumber
	+c_{16}g_{\xi}(\chi,\xi)+c_{17}g_{\xi}(\phi,\chi,\xi)= a_{22}W_{\xi\phi}(\phi,\chi) \\ 
	&&
	+a_{23}W_{\xi\phi}(\phi)+a_{25}W_{\xi\phi}(\phi,\xi)+a_{27}W_{\xi\phi}(\phi,\chi,\xi) \\ \nonumber
	&&
	+c_{32}\tilde{g}_{\phi}(\phi,\chi)+c_{33}\tilde{g}_{\phi}(\phi)+c_{35}\tilde{g}_{\phi}(\phi,\xi)+c_{37}\tilde{g}_{\phi}(\phi,\chi,\xi)\,,
	\en
	respectively. 
	
	In order to determine unique forms for $g$, $\tilde{f}$ and $\tilde{g}$, we need to establish some restrictions for constraints (\ref{sec3_eq6}), (\ref{sec3_eq10}) and (\ref{sec3_eq12}). Such restrictions may yield to different forms for the effective three-field model. Below we are going to show two different scenarios which can be generated from these restrictions. 
	
	As a first scenario, let us choose $c_{14}=c_{15}=c_{16}=c_{17}=c_{22}=c_{23}=c_{25}=c_{27}=c_{31}=c_{32}=c_{36}=c_{37}=0$ in (\ref{sec3_eq6}), (\ref{sec3_eq10}) and (\ref{sec3_eq12}), leading to
	\bn \label{sec3_eq13}
	&&
	\nonumber
	a_{11}W_{\phi\chi}(\chi)+a_{12}W_{\phi\chi}(\phi,\chi)+a_{16}W_{\phi\chi}(\chi,\xi) \\ \nonumber
	&&
	+a_{17}W_{\phi\chi}(\phi,\chi,\xi)+c_{11}g_{\chi}(\chi)+c_{12}g_{\chi}(\phi,\chi) \\ \nonumber
	&&
	= b_{2}W_{\chi\phi}(\phi,\chi)+b_{3}W_{\chi\phi}(\phi) \\ 
	&&
	+b_{5}W_{\chi\phi}(\phi,\xi)+b_{7}W_{\chi\phi}(\phi,\chi,\xi)\,,
	\en
	\bn \label{sec3_eq14}
	&&
	\nonumber
	a_{21}W_{\xi\chi}(\chi)+a_{22}W_{\xi\chi}(\phi,\chi)+a_{26}W_{\xi\chi}(\xi,\chi) \\ \nonumber
	&&
	+a_{27}W_{\xi\chi}(\phi,\chi,\xi)=b_{4}W_{\chi\xi}(\xi)+b_{5}W_{\chi\xi}(\phi,\xi) \\ \nonumber
	&&
	+b_{6}W_{\chi\xi}(\chi,\xi)+b_{7}W_{\chi\xi}(\phi,\chi,\xi) \\ 
	&&
	+c_{24}\tilde{f}_{\xi}(\xi)+c_{26}\tilde{f}_{\xi}(\chi,\xi)\,,
	\en
	\bn \label{sec3_eq15}
	&&
	\nonumber
	a_{14}W_{\phi\xi}(\xi)+a_{15}W_{\phi\xi}(\phi,\xi)+a_{16}W_{\phi\xi}(\chi,\xi) \\ \nonumber
	&&
	+a_{17}W_{\phi\xi}(\phi,\chi,\xi)=a_{22}W_{\xi\phi}(\phi,\chi)+ a_{23}W_{\xi\phi}(\phi)\\ \nonumber
	&&
	+a_{25}W_{\xi\phi}(\phi,\xi)+a_{27}W_{\xi\phi}(\phi,\chi,\xi) \\ 
	&&
	+c_{33}\tilde{g}_{\phi}(\phi)+c_{35}\tilde{g}_{\phi}(\phi,\xi)\,,
	\en
	where we still need to decide if $c_{11}=0$ or $c_{12}=0$, if $c_{24}=0$ or $c_{26}=0$, and if $c_{33}=0$ or $c_{35}=0$, in order to have unique equations for the arbitrary functions $g$, $\tilde{f}$ and $\tilde{g}$.
	
	The second scenario is built taking $c_{11}=c_{12}=c_{16}=c_{17}=c_{24}=c_{25}=c_{26}=c_{27}=c_{32}=c_{33}=c_{35}=c_{37}=0$ in (\ref{sec3_eq6}), (\ref{sec3_eq10}) and (\ref{sec3_eq12}), resulting in 
	\bn \label{sec3_eq16}
	&&
	\nonumber
	a_{11}W_{\phi\chi}(\chi)+a_{12}W_{\phi\chi}(\phi,\chi)+a_{16}W_{\phi\chi}(\chi,\xi) \\ \nonumber
	&&
	+a_{17}W_{\phi\chi}(\phi,\chi,\xi)= b_{2}W_{\chi\phi}(\phi,\chi)\\ \nonumber
	&&
	+b_{3}W_{\chi\phi}(\phi)+b_{5}W_{\chi\phi}(\phi,\xi)+b_{7}W_{\chi\phi}(\phi,\chi,\xi) \\ 
	&&
	+c_{22}\tilde{f}_{\phi}(\phi,\chi)+c_{23}\tilde{f}_{\phi}(\phi)\,,
	\en
	\bn \label{sec3_eq17}
	&&
	\nonumber
	a_{21}W_{\xi\chi}(\chi)+a_{22}W_{\xi\chi}(\phi,\chi)+a_{26}W_{\xi\chi}(\xi,\chi) \\ \nonumber
	&&
	+a_{27}W_{\xi\chi}(\phi,\chi,\xi)+c_{31}\tilde{g}_{\chi}(\chi)+c_{36}\tilde{g}_{\chi}(\chi,\xi)\\ \nonumber
	&&
	=b_{4}W_{\chi\xi}(\xi)+b_{5}W_{\chi\xi}(\phi,\xi) \\ 
	&&
	+b_{6}W_{\chi\xi}(\chi,\xi)+b_{7}W_{\chi\xi}(\phi,\chi,\xi)\,,
	\en
	\bn \label{sec3_eq18}
	&&
	\nonumber
	a_{14}W_{\phi\xi}(\xi)+a_{15}W_{\phi\xi}(\phi,\xi)+a_{16}W_{\phi\xi}(\chi,\xi) \\ \nonumber
	&&
	+a_{17}W_{\phi\xi}(\phi,\chi,\xi)+c_{14}g_{\xi}(\xi)+c_{15}g_{\xi}(\phi,\xi)\\ \nonumber
	&&
	= a_{22}W_{\xi\phi}(\phi,\chi)+a_{23}W_{\xi\phi}(\phi) \\
	&&
	+a_{25}W_{\xi\phi}(\phi,\xi)+a_{27}W_{\xi\phi}(\phi,\chi,\xi)\,.
	\en
	As in the first scenario, we have to impose if $c_{22}=0$ or $c_{33}=0$, if $c_{31}=0$ or $c_{36}$, and if $c_{14}=0$ or $c_{15}=0$, once we are looking for unique forms for $\tilde{f}$, $\tilde{g}$ and $g$. After the calculation of $g$, $\tilde{f}$ and $\tilde{g}$, we can substitute all the ingredients into  Eqs. (\ref{sec3_eq4}), (\ref{sec3_eq7}) and (\ref{sec3_eq8}) to derive $W_{\phi}$, $W_{\chi}$ and $W_{\xi}$. The next section exemplifies the applicability of our methodology and unveils new analytical three scalar field models.  

\bigskip	

\section{Examples}
\label{sec_4}
\subsection{Example I - $\phi^{\,4}$ versus $\chi^{\,4\,I}$ versus $\xi^{\,4\,I}$}
Our first example is the coupling between a $\phi^{\,4}$ model with $\chi^{\,4\,I}$, and $\xi^{\,4\,I}$, where $I$ stands for ``inverted''. The first-order differential equations for each one of these models are

\begin{figure}[h!]
\centering
\includegraphics[width=0.45\columnwidth]{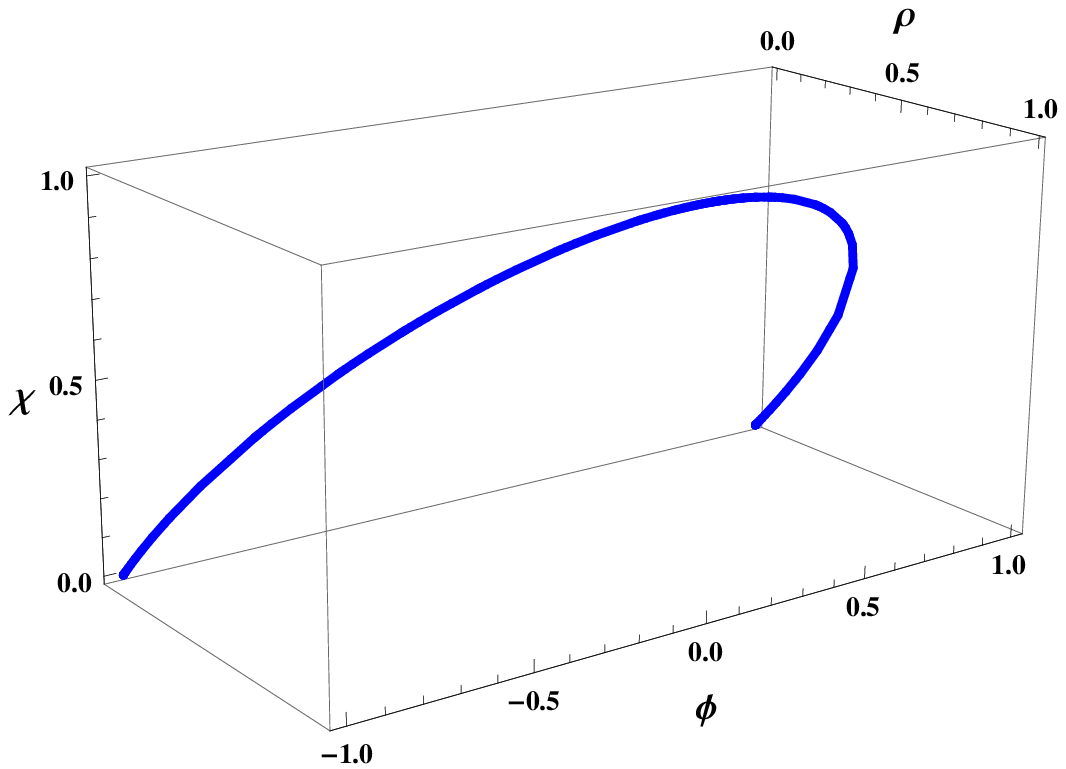} \hspace{0.2 cm} \includegraphics[width=0.45\columnwidth]{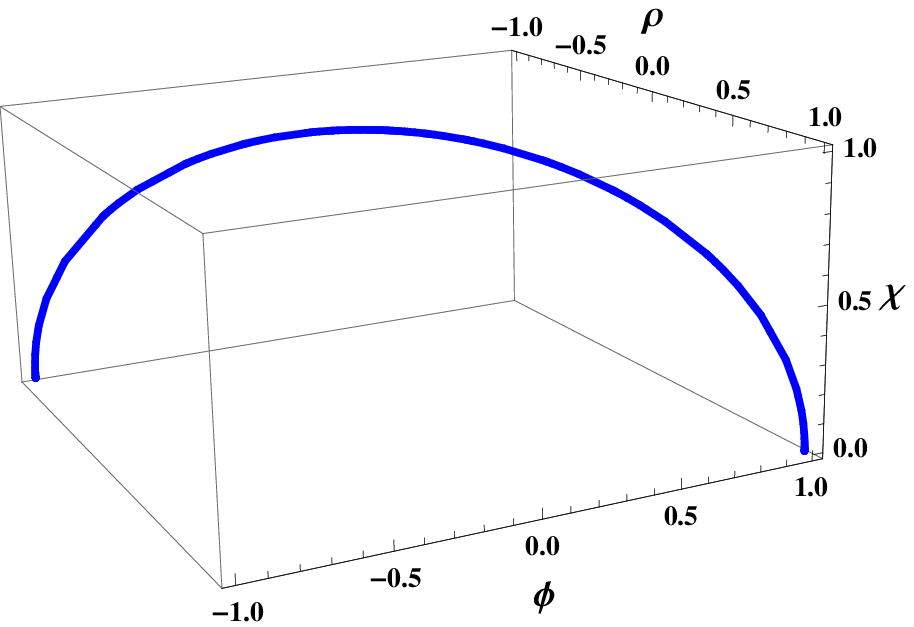}
\caption{Parametric graphics for the analytical orbits approached in examples $I$ (left panel), $II$ (right panel), all plotted with $a=b=1$.}
\label{fig1}
\end{figure}

\ben \label{sec4_eq1}
&& \nonumber
\phi^{\,\prime}=W_{\phi}(\phi)=a(1-\phi^{2})\,;\\ \nonumber
&&
\chi^{\,\prime}(\chi)= W_{\chi}(\chi)=-a\chi\sqrt{1-\frac{\chi^{2}}{b^{2}}}\,; \\
&&
\xi^{\,\prime}=W_{\xi}(\xi)=-a\xi\sqrt{1-\frac{\xi^{2}}{b^{2}}},
\een
with $a$ and $b$ real constants and whose solutions are 
\be \label{sec4_eq2}
\phi=\tanh(ax)\,;\qquad\chi=b\,\mbox{sech}(ax)\,;\qquad \xi=b\,\mbox{sech}(ax)\,.
\ee
The deformation functions (as well as their inverse functions), connecting the previous models have the forms
\be \label{sec4_eq3}
\phi=f_1(\chi)=\sqrt{1-\frac{\chi^{2}}{b^{2}}}\,; \qquad \chi=b\sqrt{1-\phi^{2}}\,;
\ee
\be\label{sec4_eq4}
\phi=f_2(\xi)=\sqrt{1-\frac{\xi^{2}}{b^{2}}}\,; \qquad \xi=b\sqrt{1-\phi^{2}}\,;
\ee
\be \label{sec4_eq5}
\xi=f_3(\chi)=\chi\,; \qquad \chi=\xi\,,
\ee
where the previous connections establish a three-dimensional orbit between the fields, which can be viewed in the left panel of Fig. \ref{fig1}. Then, we are able to use the deformations and their inverse functions to rewrite $ W_{\phi}(\phi)$, $W_{\chi}(\chi)$ and $W_{\xi}(\xi)$ in different but equivalent expressions, as we show below

\begin{widetext}
	\bn \label{sec4_eq6}
	&&
	\nonumber
	W_{\phi}(\phi)=a(1-\phi^{2})\,; \qquad W_{\phi}(\chi)=\frac{a\chi^{2}}{b^{2}},\qquad W_{\phi}(\phi,\chi)=a\left(1-\phi\sqrt{1-\frac{\chi^{2}}{b^{2}}}\right)\,;\\
	&&
	W_{\phi}(\xi)=\frac{a\xi^{2}}{b^{2}}\,;\qquad W_{\phi}(\phi,\xi)=a\left(1-\phi\sqrt{1-\frac{\xi^{2}}{b^{2}}}\right)\,;\\
	&&
	\nonumber
	W_{\phi}(\chi,\xi)=a\left(1-\sqrt{1-\frac{\chi^{2}}{b^{2}}}\sqrt{1-\frac{\xi^{2}}{b^{2}}}\right)\,; \qquad W_{\phi}(\phi,\chi,\xi)=a\left(1-\phi\sqrt{1-\frac{\chi \xi}{b^{2}}}\right)\,;
	\en
	\bn \label{sec4_eq7}
	&&
	\nonumber
	W_{\chi}(\chi)=-a\chi \sqrt{1-\frac{\chi^{2}}{b^{2}}}\,;\qquad W_{\chi}(\phi)=-ab\phi \sqrt{1-\phi^{2}}\,; \qquad W_{\chi}(\phi,\chi)=-a\chi\phi\,;\\
	&&
	W_{\chi}(\xi)=-a\xi \sqrt{1-\frac{\xi^{2}}{b^{2}}}\,;\qquad W_{\chi}(\chi,\xi)=-a\xi \sqrt{1-\frac{\chi^{2}}{b^{2}}}\,;\\
	&&
	\nonumber
	W_{\chi}(\phi,\xi)=-ab \sqrt{1-\phi^{2}} \sqrt{1-\frac{\xi^{2}}{b^{2}}}\,;\qquad W_{\chi}(\phi,\chi,\xi)=-ab \sqrt{1-\phi^{2}} \sqrt{1-\frac{\chi\xi}{b^{2}}}\,;
	\en
	and
	\bn \label{sec4_eq8}
	&&
	\nonumber
	W_{\xi}(\xi)=-a\xi \sqrt{1-\frac{\xi^{2}}{b^{2}}}\,;\qquad W_{\xi}(\phi)=-ab\phi \sqrt{1-\phi^{2}}\,; \qquad W_{\xi}(\phi,\xi)=-a\xi\phi\,; \\
	&&
	W_{\xi}(\chi)=-a\chi\sqrt{1-\frac{\chi^{2}}{b^{2}}}\,;\qquad W_{\xi}(\chi,\xi)=-a\chi \sqrt{1-\frac{\xi^{2}}{b^{2}}}\,;\\
	&&
	\nonumber
	W_{\xi}(\phi,\chi)=-ab\sqrt{1-\phi^{2}} \sqrt{1-\frac{\chi^{2}}{b^{2}}}\,;\qquad W_{\xi}(\phi,\chi,\xi)=-ab \sqrt{1-\phi^{2}} \sqrt{1-\frac{\xi\chi}{b^{2}}}\,.
	\en
\end{widetext}

The next step is to use these ingredients to derive the connection functions $g$, $\tilde{f}$ and $\tilde{g}$. In order to do it, let us consider the two possible scenarios that we pointed out in the last section, then, we can compare the similarities or the differences between these approaches.

\subsection*{First approach}

As we are interested in polynomial potentials, we are going to choose $ a_{12}=a_{15}=a_{16}=a_{17}=b_{1}=b_{3}=b_{4}=b_{5}=b_{6}=b_{7}=a_{21}=a_{22}=a_{23}=a_{24}=a_{26}=a_{27}=0$, in order to avoid rational exponents. Such choices result in the constraints $a_{11}+a_{13}+a_{14}=1$, $b_{2}=1 $ and $a_{25}=1$. Moreover, substituting these ingredients in (\ref{sec3_eq13}), we have
\be \label{sec4_eq9}
a_{11}\frac{2a\chi}{b^{2}}+c_{11}\,g_{\chi}(\chi)+c_{12}\,g_{\chi}(\phi,\chi)=-a\chi.
\ee
We choose $c_{12}=0$ as we want a unique equation to determine $g$. So, integrating the last in respect to $\chi$ we find
\be \label{sec4_eq10}
c_{11}\,g(\chi)=-a\chi^{2}\,\left(\frac{1}{2}+\frac{a_{11}}{b^{2}}\right).
\ee

The previous choice implies in $c_{11}=-c_{13}$, and the deformation function allows us to rewrite $g(\chi)$ as
\be \label{sec4_eq11}
c_{11}\,g(\phi)=-a(1-\phi^{2})\,\left(\frac{b^{2}}{2}+ a_{11} \right)\,.
\ee
Moreover, substituting the set of constraints in Eqs. (\ref{sec3_eq14}) and (\ref{sec3_eq15}) we can observe that $\tilde{f}=0$, and that
\be \label{sec4_eq12}
c_{35}\,\tilde{g}_{\phi}(\phi,\xi)=a_{14}\frac{2a\xi}{b^{2}}+a\xi\,,
\ee
for $c_{\,33}=0$. Again, we would like to avoid rational exponents in our effective model, so we are going to take $a_{14}=-b^{2}/2$, which means that $\tilde{g}=0$. Then,  putting $g$, $\tilde{f}$, $\tilde{g}$, (\ref{sec4_eq6}), (\ref{sec4_eq7}) and (\ref{sec4_eq8}) back into Eqs. (\ref{sec3_eq4}), (\ref{sec3_eq7}) and (\ref{sec3_eq8}) yields 
\ben \label{sec4_eq13}
&& \nonumber
W_{\,\phi}=-\frac{a}{2}\,\left(\chi^{\,2}+\xi^{\,2}\right)+a\,\left(1+b^2\right)\,\left(1-\phi^{\,2}\right)\,; \\ 
&&
W_{\,\chi}=-a\chi\phi\,; \qquad W_{\,\xi}=-a\phi\xi.
\een

Now, let us integrate the results presented in (\ref{sec4_eq13}) in respect to $\phi$, $\chi$ and $\xi$, respectively. Such a procedure results in the effective three-field superpotential
\be \label{sec4_eq14}
W=a\left(1+b^{2}\right)\,\left(\phi-\frac{\phi^{3}}{3}\right)-\frac{a}{2}\,\phi\,\left(\chi^{2}+\xi^{2}\right).
\ee

A very interesting case occurs when constants $a$ and $b$ are
\be \label{sec4_eq15}
a=2r\,; \qquad \qquad b=\sqrt{\frac{1}{2r}-1}\,,
\ee
with $ r\in (0,1/2)$, leading us to
\be \label{sec4_eq16}
W=\left(\phi-\frac{\phi^{3}}{3}\right)-r\,\phi\,\left(\chi^{2}+\xi^{2}\right),
\ee
whose analytical solutions have the forms
\ben \label{sec4_eq17}
&& \nonumber
\phi=\tanh(2rx)\,; \qquad\chi=\sqrt{\frac{1}{2r}-1}\,\mbox{sech}(2rx)\,;\\ 
&&
\xi=\sqrt{\frac{1}{2r}-1}\,\mbox{sech}(2rx)\,.
\een
This is the three-field version for the BNRT model \cite{bnrt} presented in \cite{blochbranes} and we highlight that it is the first time that such a model is constructed via one scalar field systems. 

\subsection*{Second approach}

In the second scenario we also need to consider $a_{12}=a_{15}=a_{16}=a_{17}=b_{1}=b_{3}=b_{4}=b_{5}=b_{6}=b_{7}=a_{21}=a_{22}=a_{23}=a_{24}=a_{26}=a_{27}=0 $ once we would like to avoid rational powers in the effective potential (or superpotential). Consequently we keep the constraints $a_{11}+a_{13}+a_{14}=1$, $b_{2}=1$ and $a_{25}=1$. Now,  putting these ingredients back into Eq. (\ref{sec3_eq16}) we obtain
\be \label{sec4_eq18}
c_{22}\,\tilde{f}_{\phi}(\phi,\chi)=a\chi+a_{11}\frac{2a\chi}{b^{2}}\,,
\ee
with $c_{23}=0$. Besides, here we must take $a_{11}=-b^{2}/2$ to avoid rational exponents in our potential, which means $\tilde{f}=0$. Looking at Eq. (\ref{sec3_eq17}) we directly see that the previous constraints impose $\tilde{g}=0$. Moreover, Eq. (\ref{sec3_eq18}) unveils that
\be \label{sec4_eq19}
a_{14}\frac{2\xi a}{b^{2}}+c_{14}\,g_{\xi}(\xi)=-a\xi,
\ee
for $c_{15}=0$  ($c_{14}=-c_{13} $). So, integrating it in respect to $\xi$ yields
\be \label{sec4_eq20}
c_{14}\,g(\xi)=\frac{a\xi^{2}}{2b^{2}}\,\left(b^{2}+2a_{14}\right).
\ee
Then, we are able to use the deformation function to rewrite the last equation as 
\be \label{sec4_eq21}
c_{14}\,g(\phi)=\frac{a(1-\phi^{2})}{2}\,\left(b^{2}+2a_{14}\right)\,.
\ee

All these ingredients enable us to use (\ref{sec3_eq4}), (\ref{sec3_eq7}) and (\ref{sec3_eq8}) to find
\ben \label{sec4_eq22}
&& \nonumber
W_{\phi}=-\frac{a}{2}\,\left(\chi^{\,2}+\xi^{\,2}\right)+a\,\left(1+b^2\right)\,\left(1-\phi^{\,2}\right)\,; \\
&&
W_{\chi}=-a\chi\phi\,; \qquad W_{\xi}=-a\phi\xi\,.
\een
Now, integrating the last equations in respect to their fields, it is possible to derive
\be \label{sec4_eq23}
W=a\left(1+b^{2}\right)\,\left(\phi-\frac{\phi^{3}}{3}\right)-\frac{a}{2}\,\phi\,\left(\chi^{2}+\xi^{2}\right)\,
\ee
as the effective three-field superpotential for this scenario. We note that Eqs. (\ref{sec4_eq23}) and (\ref{sec4_eq14}) are the same. Therefore, as far as we verified, there are no differences in the final form of the superpotential $W$ if one chooses the first or the second approach or even other possible scenarios for Eqs. (\ref{sec3_eq5}), (\ref{sec3_eq9}) and (\ref{sec3_eq11}). Based on this, we are going to consider just the first scenario approach for the next examples.  

\bigskip

\subsection{Example II - $\phi^{\,4}$ versus $\chi^{\,4\,I}$ versus $\xi^{\,4}$}

In this example we work with a combination of $ \phi^{4}$, with an inverted $\chi^{\,4\,I}$ and with $\xi^{4}$. The first-order differential equations and their solutions are
\ben \label{sec4_eq24}
&&\nonumber
\phi^{\,\prime}=W_{\phi}=a(1-\phi^{2})\,; \,\,\, \chi^{\,\prime}=W_{\chi}=-a\chi\sqrt{1-\frac{\chi^{2}}{b^{2}}}\,; \\ 
&&
\xi^{\,\prime}=W_{\xi}=a(1-\xi^{2})\,;
\een
\be \label{sec4_eq25}
\phi=\tanh(ax)\,;\qquad \chi=b\,\mbox{sech}(ax)\,; \qquad \xi=\tanh(ax)\,,
\ee
respectively. So, the deformation functions (and their inverse functions) 
\be \label{sec4_eq26}
\phi=\xi\,; \qquad \xi=\phi\,;
\ee
\be \label{sec4_eq27}
\phi=\sqrt{1-\frac{\chi^{2}}{b^{2}}}\,; \qquad \chi=b\sqrt{1-\phi^{2}}\,;
\ee
\be \label{sec4_eq28}
\xi=\sqrt{1-\frac{\chi^{2}}{b^{2}}}\,; \qquad \chi=b\sqrt{1-\xi^{2}}\,
\ee
connect these three one-field models. The three-dimensional parametric plot for the effective orbit connected by these solutions is shown in the right panel of Fig. \ref{fig1}.  
\begin{widetext}
	As in the previous examples, we can use these ingredients to rewrite $W_{\phi}(\phi)$, $W_{\chi}(\chi)$ and $W_{\xi}(\xi)$ in the following equivalent forms
	\bn \label{sec4_eq29}
	&&
	\nonumber
	W_{\phi}(\phi)=a(1-\phi^{2})\,;\qquad W_{\phi}(\chi)=\frac{a\chi^{2}}{b^{2}}\,; \qquad W_{\phi}(\phi,\chi)=\frac{a\chi}{b}\sqrt{1-\phi^{2}}\,; \qquad W_{\phi}(\xi)=a(1-\xi^{2})\,;\\
	&&
	W_{\phi}(\phi,\xi)=a(1-\xi\phi)\,;\qquad W_{\phi}(\chi,\xi)=\frac{a\chi}{b}\sqrt{1-\xi^{2}}\,; \qquad W_{\phi}(\phi,\chi,\xi)=\frac{a\chi}{b}\sqrt{1-\xi\phi}\,;
	\en
	\bn \label{sec4_eq30}
	&&
	\nonumber
	W_{\chi}(\chi)=-a\chi \sqrt{1-\frac{\chi^{2}}{b^{2}}}\,; \qquad W_{\chi}(\phi)=-ab\phi \sqrt{1-\phi^{2}}\,; \qquad W_{\chi}(\phi,\chi)=-a\chi\phi\,; \\
	&&
	\nonumber
	W_{\chi}(\xi)=-ab\xi \sqrt{1-\xi^{2}}\,; \qquad W_{\chi}(\chi,\xi)=-a\xi\chi\,; \qquad W_{\chi}(\phi,\xi)=-ab\phi \sqrt{1-\xi^{2}}\,;\\
	&&
	W_{\chi}(\phi,\chi,\xi)=-ab \sqrt{1-\frac{\chi^{2}}{b^{2}}}\sqrt{1-\xi\phi}\,;
	\en
	\bn \label{sec4_eq31}
	&&
	\nonumber
	W_{\xi}(\xi)=a(1-\xi^{2})\,; \qquad W_{\xi}(\phi)=a(1-\phi^{2})\,; \qquad W_{\xi}(\phi,\xi)=a(1-\xi\phi)\,; \\
	&&
	\nonumber
	W_{\xi}(\chi)=\frac{a\chi^{2}}{b^{2}}\,; \qquad W_{\xi}(\chi,\xi)=\frac{a\chi}{b}\sqrt{1-\xi^{2}}\,; \qquad W_{\xi}(\phi,\chi)=\frac{a\chi}{b}\sqrt{1-\phi^{2}}\,; \\
	&&
	W_{\xi}(\phi,\chi,\xi)=\frac{a\chi}{b}\sqrt{1-\xi\phi}\,.
	\en
\end{widetext}
Since we would like to avoid rational powers in our polynomial potential, we need to take $a_{12}=a_{16}=a_{17}=b_{1}=b_{3}=b_{4}=b_{5}=b_{7}=a_{22}=a_{26}=a_{27}=0$. Besides, we choose to work with $c_{12}=0$, leading to the constraints $a_{11}+a_{13}+a_{14}+a_{15}=1$, $b_{2}+b_{6}=1$, $a_{23}+a_{24}+a_{25}=1$ and $c_{11}=-c_{13}$. Therefore, Eq.(\ref{sec3_eq13}) yields
\be \label{sec4_eq32}
c_{11}\,g_{\chi}(\chi)=-b_{2}a\chi-a_{11}\frac{2a\chi}{b^{2}},
\ee
whose integration in respect to $\chi$ gives
\be \label{sec4_eq33}
c_{11}\,g(\chi)=-\frac{a\chi^{2}}{2b^{2}}\,\left(b_{2}b^{2}+2a_{11}\right)\,.
\ee

Now, we are able to use the deformation function to rewrite the last equation as
\be \label{sec4_eq34}
c_{11}\,g(\phi)=-\frac{a(1-\phi^{2})}{2}\,\left(b_{2}b^{2}+2a_{11}\right).
\ee
Besides, we need to impose that $a_{21}=b_{6}=0$ to avoid rational exponents in the effective potential, which means that $\tilde{f}=0$ (see Eq. (\ref{sec3_eq14})). Moreover, taking $c_{33}=0$ in Eq. (\ref{sec3_eq15}) we directly determine that
\be \label{sec4_eq35}
-2a_{14}a\xi-a_{15}a\phi=-2a_{23}a\phi-a_{25}a\xi+c_{35}\,\tilde{g}_{\phi}(\phi,\xi)\,,
\ee
which can be integrated in respect to $\phi$, giving rise to
\be \label{sec4_eq36}
c_{35}\,\tilde{g}(\phi,\xi)=a\,\left(a_{25}-2\,a_{14}\right)\,\xi\,\phi-\frac{a}{2}\,\left(a_{15}-2\,a_{23}\right)\,\phi^{\,2}\,.
\ee
This last expression is also represented as
\be \label{sec4_eq37}
c_{35}\,\tilde{g}(\xi)=a\,\left(a_{23}+a_{25}-\frac{a_{15}}{2}-2\,a_{14}\right)\,\xi^{\,2}\,,
\ee
via deformation function.

	Thus, applying all these results in Eqs. (\ref{sec3_eq4}), (\ref{sec3_eq7}) and (\ref{sec3_eq8}) we have
	\bn \label{sec4_eq38}
	&& \nonumber
	W_{\phi}=a\,a_{13}\,\left(1-\phi^{\,2}\right)+a\,a_{14}\,\left(1-\xi^{\,2}\right)+ \\ \nonumber
	&&
	+a\,a_{15}\,\left(1-\xi\,\phi\right)-\frac{a}{2}\,\chi^{2}+\frac{a}{2}\,\left(b^{\,2}+2\,a_{11}\right)\,\left(1-\phi^{\,2}\right)\,; \\
	&&
	W_{\chi}=-a\phi\chi\,; \qquad W_{\xi}=a\,\left(1-\xi^{\,2}\right)+\\ \nonumber
	&&
	-2\,a\,a_{14}\,\left(\phi-\xi\right)\,\xi-a\,\frac{a_{15}}{2}\,\left(\phi^{\,2}-\xi^{\,2}\right)\,,
	\en
	whose correspondent superpotential is
	\ben \label{sec4_eq39}
	\nonumber
	W &=& a\,\left(a_{11}+a_{13}+\frac{b^2}{2}\right)\,\left(\phi-\frac{\phi^{\,3}}{3}\right)+a\,a_{14}\,\left(1-\xi^{\,2}\right)\,\phi \\ 
	&&
	+a\,a_{15}\,\left(1-\frac{\xi}{2}\,\phi\right)\,\phi -\frac{a}{2}\,\chi^{\,2}\,\phi\\ \nonumber
	&&
	+a\,\left(\xi-\frac{\xi^{\,3}}{3}\right)+\frac{a}{3}\,\left(2\,a_{14}+\frac{a_{15}}{2}\right)\,\xi^{\,3}+\kappa\,,
	\een
	where $\kappa$ is a real integration constant, and $a_{11}+a_{13}+a_{14}+a_{15}=1$.
	This effective three-field superpotential is a new model in the literature and has Eq. (\ref{sec4_eq25}) as the analytical solutions of its equations of motion. 
	
	 An interesting feature of this new model is that it represents two domain walls with an internal structure. Models like this emerged before in the work of Shifmann {\it et al.}, where the authors investigated localization of gauge fields inside of domain walls \cite{shifmann_2004}. Besides, the model here derived complements the discussions presented by Bazeia {\it et al. } in \cite{blw_2002}, where the authors worked with an analytical model composed by one domain wall with an internal structure formed by two other fields. Another special aspect about both, our model and the one from \cite{blw_2002}, is that they represent a natural bridge for the four-field model introduced by Callen and Volkas \cite{callen_2013}, which has two domains walls plus an internal structure composed by two other fields. 
	
	The resultant potential can be derived combining $(\ref{sec4_eq38})$ with $(\ref{sec2_eq9})$, and one can see that it has $Z_2$ symmetry ($\phi\rightarrow -\phi, \chi \rightarrow -\chi, \xi \rightarrow -\xi$), securing the stability of this topological configuration of fields, \cite{callen_2013}. The $Z_2$ symmetry implies that our model has the same features of the Dirichlet domain walls introduced by Carroll and Trodden \cite{carroll_1998}. Such a symmetry also matches with the behavior of the model studied by Bazeia {\it et al. } in \cite{blw_2002}, moreover, we can follow the ideas presented in the mentioned work to discuss the physical features of the internal structure of our model. 
	
	Firstly, from $(\ref{sec4_eq38})$, we can see that inside both walls we have
	\ben
	&&
	W_\phi(0,\chi,0)=a\,\left(1+\frac{b^{\,2}}{2}-\frac{\chi^{\,2}}{2}\right)\,; \nonumber \\
	&&
	W_\chi(0,\chi,0)=0\,; \qquad W_\xi(0,\chi,0)=a^{\,2}	\,,
	\een
	therefore, $\chi=\sqrt{b^{\,2}+2}$ in order to maximize $V$ at this region. Furthermore, the projections of $V$ inside and outside both walls are
	\be
	V(0,\chi,0)=\frac{a^{\,2}}{8}\,\left[4+(2+b^{\,2}-\chi^{\,2})^{\,2}\right]\,;
	\ee
	\be
	V(\pm1,\chi,\pm1)=\frac{a^{\,2}}{8}\,\chi^{\,2}\,\left(4+\chi^{\,2}\right)\,,
	\ee
	respectively. So, these previous equations yield to the following masses for the scalar meson related with the internal structure
	\ben
	&&
	m_{in}^{\,2}=V_{\,\chi\,\chi}(0,\sqrt{2+b^{\,2}},0)=a^{\,2}\,(b^{\,2}+2)\,; \nonumber \\
	&&
	m_{out}^{\,2}=V_{\,\chi\,\chi}(\pm1,0,\pm1)=a^{\,2}\,; \qquad V_{\,\chi\,\chi}=\frac{\partial^{\,2}V}{\partial\,\chi^{\,2}}\,,
	\een
	leading to the ratio
	\be
\frac{m_{in}^{\,2}}{m_{out}^{\,2}}=b^{\,2}+2\,.
	\ee
The last ratio unveils that the scalar meson prefers to live outside the domain walls.

\section{Analytical three-field cosmological model}
\label{sec_5}
A potential application of our method consists in the study of cosmological models, where the Einstein-Hilbert Lagrangian is coupled with a three scalar field Lagrangi\-an density. Such an approach can be used to describe different dynamical stages that the Universe has passed through. In order to implement this discussion, let us consider the following action
\bn \label{sec5_eq1}
&& \nonumber
S=\int\,d^{\,4}\,x\,\sqrt{-g}\,\left[-\frac{R}{4}+{\cal L}(\phi_i,\partial_{\,\mu}\phi_i)\right]; \\ 
&&
{\cal L}=\sum_{i=1}^{\,3}\,\left[\frac{1}{2}\,\partial_{\,\mu}\,\phi_{\,i}\,\partial^{\,\mu}\,\phi_i-V(\phi_1,\phi_2,\phi_3)\right]\,,
\en
with $i=1,2,3$, $\phi_1=\phi(t)$, $\phi_2=\chi(t)$, $\phi_3=\xi(t)$, $4\,\pi\,G=1$, $c=1$ and signature $(+,-,-,-)$. 

The minimization of the previous action in respect to the metric yields the Einstein equations
\be \label{sec5_eq2}
R_{\,\mu\,\nu}-\frac{1}{2}\,g_{\,\mu\,\nu}\,R=2\,T_{\,\mu\,\nu}\,,
\ee
where $T_{\,\mu\,\nu}$ is the energy-momentum tensor whose explicit form is
\be \label{sec5_eq3}
T_{\,\mu\,\nu}=2\,\frac{\partial\,{\cal L}}{\partial\,g^{\,\mu\,\nu}}-g_{\,\mu\,\nu}\,{\cal L}\,,
\ee
and has components $(\rho,-p,-p,-p)$, where $\rho$ and $p$ are the density and the pressure related with the scalar field model. From the previous equation, we are able to compute
\be \label{sec5_eq4}
\rho=\sum_{i=1}^{3}\,\frac{\dot{\phi}_{\,i}^{\,2}}{2}+V\,; \qquad
p=\sum_{i=1}^{3}\,\frac{\dot{\phi}_{\,i}^{\,2}}{2}-V\,. \\ 
\ee
Moreover, a flat Friedmann-Robertson-Walker metric leads us to the Friedmann equations
\be \label{sec5_eq5}
H^{\,2}=\frac{2}{3}\,\rho\,; \qquad 
\dot{H}+H^{\,2} =-\frac{1}{3}\,\left(\rho+3\,p\right)\,,
\ee
where $H$ is the Hubble parameter. 

Equations (\ref{sec5_eq5}) can be rewritten as 
\ben \label{sec5_eq6}
&& \nonumber
H^{\,2}=\frac{1}{3}\,\left(\dot{\phi}^{\,2}+\dot{\chi}^{\,2}+\dot{\xi}^{\,2}+2\,V\right)\,; \\
&&
\dot{H}=-\left(\dot{\phi}^{\,2}+\dot{\chi}^{\,2}+\dot{\xi}^{\,2}\right)\,.
\een
Besides the Hubble parameter, another interesting quantity to analyze is the so-called equation of state (EoS) parameter, which is the ratio between pressure and density of the observed Universe, i.e.,
\be \label{sec5_eq6_1}
\omega=\frac{p}{\rho}\,.
\ee

A first-order formalism is implemented by defining

\be \label{sec5_eq7}
H=-W(\phi,\chi,\xi)\,,
\ee
which means that

\be \label{sec5_eq8}
\dot{H}=-W_{\,\phi}\,\dot{\phi}-W_{\,\chi}\,\dot{\chi}-W_{\,\xi}\,\dot{\xi}\,.
\ee 
By substituting $H$ and $\dot{H}$ into (\ref{sec5_eq6}), we find the first-order differential equations
\be \label{sec5_eq9}
\dot{\phi}=W_{\,\phi}\,; \,\, \dot{\chi}=W_{\,\chi}\,;\,\,\dot{\xi}=W_{\,\xi}\,,
\ee 
and the potential
\be \label{sec5_eq10}
V=\frac{3}{2}\,W^{\,2}-\frac{1}{2}\,\left(W_{\,\phi}^{\,2}+W_{\,\chi}^{\,2}+W_{\,\xi}^{\,2}\right)\,.
\ee 
Moreover, the minimization process of the action (\ref{sec5_eq1}) in respect to the fields yields the equations of motion
\be \label{sec5_eq11}
\ddot{\phi}_{\,i}+3\,H\,\dot{\phi}_{\,i}+V_{\phi_{\,i}}=0\,; \,\, i=1,\,2,\,3,
\ee
which need to be satisfied by the solutions of (\ref{sec5_eq9}).

\begin{figure}[ht!]
\centering
\includegraphics[width=0.75\columnwidth]{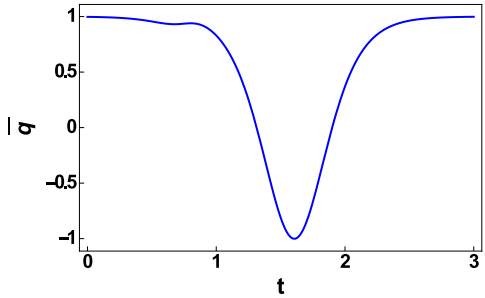}
\caption{Time evolution of the analytical acceleration parameter derived from a three-field model.}
\label{fig2}
\end{figure}

\begin{figure}[ht!]
\centering
\includegraphics[width=0.75\columnwidth]{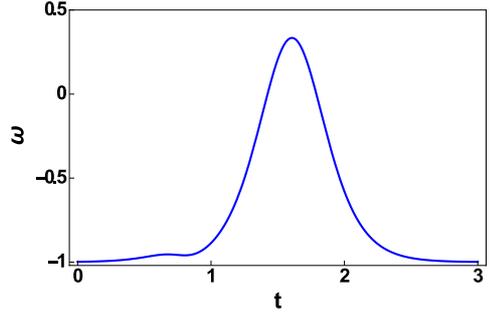}
\caption{Time evolution of the analytical EoS parameter derived from a three-field model.}
\label{fig3}
\end{figure}

Let us apply our new model, introduced in (\ref{sec4_eq39}), in such a cosmological scenario. The solutions, which are going to satisfy the first-order equations (\ref{sec5_eq9}),  are
\be \label{sec5_eq12}
\phi(t)=\xi(t)=\tanh\,\left(a\,t+\tau\right)\,; \,\, \chi(t)=b\,\mbox{sech}\,\left(a\,t+\tau\right)\,,
\ee
where $a$, $b$ and $\tau$ are real constants. 

These previous solutions together with (\ref{sec4_eq39}) allow us to determine the Hubble parameter
\ben \label{sec5_eq13}
&& \nonumber
H(t)=\frac{1}{6} \bigg[a \left(b^2+4\right) \tanh ^3(a t+\tau )+3 a \tanh (a t+\tau )  \\
&&
\times\,\left(b^2 \text{sech}^2(a t+\tau )-b^2-4\right)-6 \kappa \bigg]\,.
\een
 From (\ref{sec5_eq13}), we are able to plot the acceleration parameter $\bar{q}=H^{-1}(\ddot{\alpha}/\dot{\alpha})$, defined so that positive values of $\bar{q}$ indicate an accelerated expansion of the Universe, while negative values indicate a decelerated expansion. The evolution of $\bar{q}$ in time is shown in Fig. \ref{fig2}, where we assumed $a=3$, $b=3.5$,  $\kappa=-16.98$ and $\tau =-2.5$. It is relevant to say that such parameters were chosen in order to derive a viable cosmological scenario. However, the values adopted by $a$, $b$, $\kappa$, and $\tau$ are not extremely constrained, or in another words, small variations of these values do not change too much the physical aspects of the cosmological parameters.

The previous results enable us to determine $V$ in Equation (\ref{sec5_eq10}) and the EoS (\ref{sec5_eq6_1}) respectively as
\begin{widetext}
\ben \label{sec5_eq14}
&& \nonumber
V=\frac{1}{24} \bigg[-3 \, a^2 \bigg(-2\, \phi ^2 \,(a_{14}+a_{15}-1)+2\, a_{14}\, \xi ^2+2\, a_{15}\, \xi \, \phi +b^2 \,\left(\phi ^2-1\right)+\chi ^2-2\bigg)^2-3 a^2\, \bigg( (4\, a_{14}+a_{15}-2)\,\xi ^2 \\ \nonumber
&&
-4\, a_{14}\, \xi \, \phi -a_{15}\, \phi ^2+2\bigg)^2-12\, a^2\, \chi ^2 \,\phi ^2+\bigg(a\, \left[\phi\,  \left(2\, \phi ^2 (a_{14}+a_{15}-1)-b^2\,\left(\phi ^2-3\right)-3\, \chi ^2+6\right)\right.
 \\
&&
\left.+\xi ^3\, (4 \,a_{14}+a_{15}-2)-6 \,a_{14}\, \xi ^2 \,\phi +\xi\,  \left(6-3\, a_{15}\, \phi ^2\right)\right]+6\, \kappa \bigg)^2\bigg]\,;
\een
\ben \label{sec5_eq15}
&&
\omega=-\bigg\{a^2\, b^4 \cosh (6 (a t+\tau ))+8\, a^2\, b^2\, \cosh (6 (a t+\tau ))+3\, \left(a^2\, \left(5\, b^4-24\,b^2-48\right)+45\, \kappa ^2\right) \cosh (2 (a t+\tau )) \\ \nonumber
&&
-6\, \left(a^2 \left(b^4+4\, b^2-16\right)-9\, \kappa ^2\right) \cosh (4 (a t+\tau ))-2\, a^2\, \left(5\, b^4-44\, b^2+176\right)+16 \,a^2 \cosh (6 (a t+\tau )) \\ \nonumber
&&
-18\, a\, b^2 \,\kappa\,  \sinh (2 (a t+\tau ))+6\, a\, \left(b^2+4\right)\, \kappa  \,\sinh (6 (a t+\tau ))+9\, \kappa ^2\, \cosh (6 (a t+\tau ))+216\, a\, \kappa\,  \sinh (2 (a t+\tau ))+90\, \kappa ^2 \\ \nonumber
&&
+144\, a\, \kappa\,  \sinh (4 (a t+\tau ))\bigg\}\,\left\{2\, \left(a \,\left(b^2+4\right)\, \sinh (3 (a t+\tau ))-3\, a\, \left(b^2-4\right)\, \sinh (a t+\tau )+12\, \kappa\,  \cosh ^3(a t+\tau )\right)^2\right\}^{\,-1}\,.
\een
\end{widetext}
 The behavior of the EoS parameter can be visualized in Fig. \ref{fig3}. We can point that the graphics presented in Figs \ref{fig2} and \ref{fig3} agree it other, unveiling two inflationary eras for the early and late times, besides a decelerated era between the two stages of acceleration. We also see that $\omega \approx -1$ in both of these accelerated eras, simulating a dark energy domination as time passes by \cite{planck}. Furthermore, we can use $V$ (\ref{sec5_eq14}) together with $W$ (\ref{sec4_eq39}) and the solutions (\ref{sec5_eq12}) to prove that the equations of motion (\ref{sec5_eq11}) are satisfied.

\section{Conclusion}
\label{sec_6}

A mechanism to generate new models with three scalar fields was presented in this work. We started the method coupling three analytical one-field models via deformation procedure, introduced in \cite{bls}. The non-trivial combinations of these one-field systems unveiled effective three-field models. As a first example we were able to derive a well-known three-field version for the BNRT model \cite{blochbranes}. Another interesting feature is that the new models are automatically satisfied by the solutions of the one-field systems, corroborating with the results derived in \cite{bls}. 

The superpotentials here derived are all polynomial, but we also can use this methodology to build three-field models with functional potentials, like combinations involving sine-Gordon potentials, for instance. Besides, with this superpotential in hands we are able to find the total energy related with the correspondent defects solutions, as well as, the potential $V(\phi,\chi,\xi)$.

 The mechanism has shown to provide an interesting cosmological scenario, able to predict two accelerated eras, including a latte time accelerated one, which simulates the dark energy era. The acceleration and EoS parameters for the decelerated stages of the Universe have also been obtained, which leads us to conclude that from the deformation procedure applied to a scenario with three scalar fields, one is able to obtain a complete cosmological scenario, with the transition stages being described continuously. 
 
We believe that the method here presented can be applied in compactons-like defects \cite{bm_15}, in braneworld \cite{blochbranes,bm_15} and in another cosmological scenarios \cite{ms_14,ms_16,bp_12}. It is going to be an interesting task to observe the consequence of such new analytical three-field systems for the physical parameters like the cosmological ones or the warp-factor. Some of these ideas are under investigation and we hope to report them in near future.

\

{\bf Acknowledgments} \ D. A. Ferreira and  D. C. Vilar Neta would like to thank Capes (Brazilian agency) for financial support. P. H. R. S. Moraes would like to thank S\~ao Paulo Research Foundation (FAPESP), grant
2015/08476-0, for financial support. The authors also would like to thank both anonymous referees for their criticism, which enhanced the potential of this work.

\end{document}